\documentclass[twocolumn,showpacs,aps,prl,superscriptaddress]{revtex4}

\usepackage{graphicx}
\usepackage{dcolumn}
\usepackage{amsmath}
\usepackage{epsfig}
\usepackage{longtable}

\input babarsym

\def\pom {\ensuremath{\pm}\xspace}

\def\Imm       {\ensuremath{\Im m}}
\def\Ree       {\ensuremath{\Re e}}

\newcommand{\dd}{\text{d}}

\newcommand{\BABARPubYear}    {07}
\newcommand{\BABARPubNumber}  {009}

\newcommand{\SLACPubNumber} {12430}
\newcommand{\LANLNumber} {0704.0522}

\def\figurebox#1#2#3{%
    \def\arg{#3}%
    \ifx\arg\empty
    {\hfill\vbox{\hsize#2\hrule\hbox to #2{\vrule\hfill\vbox to #1{\hsize#2\vfill}\vrule}\hrule}\hfill}%
    \else
    {\hfill\epsfbox{#3}\hfill}%
    \fi}

\def\Kstarone   {\ensuremath{K^*(892)}\xspace}

\def\mell       {\ensuremath{\ell}\xspace}

\newcommand{\vomega}{\boldsymbol{\omega}}
\newcommand{\jchan}{{b}}
\newcommand{\ichan}{{a}}

\begin{document}

\preprint{\babar-PUB-\BABARPubYear/\BABARPubNumber} 
\preprint{SLAC-PUB-\SLACPubNumber} 

\begin{flushleft}

\babar-PUB-\BABARPubYear/\BABARPubNumber\\
SLAC-PUB-\SLACPubNumber\\
hep-ex/\LANLNumber
\end{flushleft}

\title{{\large \bf Measurement of Decay Amplitudes of $\B \to (\ccbar) \Kstar$ with an Angular Analysis, for $(\ccbar)=\jpsi$, \psitwos and \chicone} }

%
\author{B.~Aubert}
\author{M.~Bona}
\author{D.~Boutigny}
\author{Y.~Karyotakis}
\author{J.~P.~Lees}
\author{V.~Poireau}
\author{X.~Prudent}
\author{V.~Tisserand}
\author{A.~Zghiche}
\affiliation{Laboratoire de Physique des Particules, IN2P3/CNRS et Universit\'e de Savoie, F-74941 Annecy-Le-Vieux, France }
\author{J.~Garra~Tico}
\author{E.~Grauges}
\affiliation{Universitat de Barcelona, Facultat de Fisica, Departament ECM, E-08028 Barcelona, Spain }
% ...
\collaboration{The \babar\ Collaboration}
\noaffiliation

\date{\today}

\begin{abstract}
We perform the first three-dimensional measurement of the amplitudes of
$\B\to\psitwos\Kstar$ and $\B\to\chicone\Kstar$ decays and update our
previous measurement for $\B\to\jpsi\Kstar$.
We use a data sample collected with the \babar\ detector at the \pep2\
storage ring, corresponding to 232 million \BB\ pairs.
The longitudinal polarization of decays involving a $J^{PC}=1^{++}$ \chicone
meson is found to be larger than that with a $1^{--}$ $J/\psi$ or $\psi(2S)$ meson.
No direct \CP-violating charge asymmetry is observed.
\end{abstract}


\maketitle

In the context of measuring the parameters of the Unitarity Triangle
of the CKM matrix, \Bz decays to charmonium-containing final states
(\jpsi, \psitwos, \chicone)\Kstar,
defined collectively here as $B^0\rightarrow (c\bar c) K^*$,
are of interest for the precise measurement of $\sin 2 \beta$, where
$\beta\equiv \arg[-V_{cd}V_{cb}^*/V_{td}V_{tb}^*]$, in a similar way as
for $\B^0 \to \jpsi \Kz$. Furthermore, the $\jpsi\Kstar$ channel allows
the measurement of $\cos 2 \beta$ \cite{Aubert:2004cp}.

For the modes considered in this paper, the final state 
consists of two spin-1 mesons, leading to three possible
values of the total angular momentum with different \CP\ 
eigenvalues ($L = 1$ is odd, while $L=0,2$ are even). The 
different contributions must be taken into account in the
measurement of $\sin2\beta$. The amplitude for longitudinal 
polarization of the two spin-1 mesons is $A_0$. 
There are two amplitudes for polarizations of the mesons transverse to
the decay axis, here expressed in the transversity basis
\cite{Dunietz:1990cj}: $A_\parallel $ for parallel polarization and
$A_\perp$ for their perpendicular polarization.
Only the relative amplitudes are measured, so that 
$|A_0|^2 + |A_\parallel |^2 + |A_\perp|^2 = 1$.
Previous measurements by the 
CLEO \cite{Jessop:1997jk},
CDF \cite{Affolder:2000ec},
\babar\ \cite{Aubert:2004cp}
and Belle \cite{Itoh:2005ks}
collaborations for the $\B\to\jpsi\Kstar$ channels are all compatible with
each other, and with a \CP-odd intensity fraction $|A_\perp|^2$ close to
0.2.

Factorization predicts 
that the phases of the transversity decay amplitudes are the same. 
\babar\ has observed \cite{Aubert:2004cp,Aubert:2001pe}
a significant departure from this prediction.

Precise measurements of the branching fractions of $B\rightarrow (c\bar c) K^*$ decays are
now available \cite{Aubert:2004rz} to test the theoretical description
of the non-factorizable contributions \cite{Chen:2005ht}, but
polarization measurements are also needed.
In particular, measurements for \psitwos and \chicone, compared to
that of \jpsi, would discriminate the mass dependence from the quantum
number dependence. CLEO has measured the longitudinal polarization of 
 \B\to\psitwos\Kstar decays to be $|A_0|^2 =0.45 \pm 0.11 \pm 0.04 $ 
\cite{Richichi:2000ca}.
Belle has studied \B\to\chicone\Kstar
decays and obtained $|A_0|^2 =0.87 \pm 0.09 \pm 0.07$ \cite{Soni:2005fw}.

$B\rightarrow (c\bar c) K^{(*)}$ decays provide a clean environment for the
measurement of the CKM angle $\beta$ because one tree amplitude dominates
the decay. 
Very small direct \CP-violating charge asymmetries are
expected in these decays, and
no such signal has been found \cite{Aubert:2004rz}.
While more than one amplitude with different strong and weak phases
are needed to create a charge asymmetry in a simple branching fraction
measurement, London {\em et al.} have suggested \cite{London:2000zi}
that an angular analysis of vector-vector decays can detect charge
asymmetries even in the case of vanishing strong phase difference.
Belle has looked for, and not found, such a signal \cite{Itoh:2005ks}.

In this paper we present the amplitude measurement of charged and
neutral $\B \to (\ccbar)\Kstar$ using a selection similar to that of
Ref.~\cite{Aubert:2004rz}, and a fitting method similar to that of
Ref.~\cite{Aubert:2004cp}.
We use the notation $\psi$ for the $1^{--}$ states \jpsi and \psitwos. 
$\psi$ (\chicone) candidates are reconstructed in their decays to $\ellp\ellm$ ($\jpsi\g$),
where \mell represents an electron or a muon.
Decays to the flavor eigenstates $\Kstarz\to\Kpm\pimp$,
$\Kstarpm\to\KS\pipm$ and $\Kstarpm\to\Kpm\piz$ are used.
The relative strong phases are known to have a two-fold
ambiguity when measured in an angular analysis alone.
In contrast to earlier publications
\cite{Jessop:1997jk,Affolder:2000ec,Aubert:2001pe} we use here the set
of phases predicted in Ref.~\cite{Suzuki:2001za}, with arguments
based on the conservation of the $s$-quark helicity in the decay of
the $b$ quark.
We have confirmed experimentally this prediction through the study of
the variation with $K\pi$ invariant mass of the phase difference 
between the \Kstarone amplitude and a non-resonant $K\pi$ $S$-wave amplitude
\cite{Aubert:2004cp}.

The data were collected with the \babar\ detector at the \pep2\ asymmetric
\epem storage ring, 
and correspond to an integrated luminosity of about 209 \invfb at the
center-of-mass energy near the \FourS mass.
The \babar\ detector is described in detail elsewhere~\cite{detector}. 
Charged-particle tracking is provided by a five-layer silicon
vertex tracker (SVT) and a 40-layer drift chamber (DCH). 
For charged-particle identification (PID), ionization energy loss in
the DCH and SVT, and Cherenkov radiation detected in a ring-imaging
device (DIRC) are used.
Photons are identified by the electromagnetic calorimeter
(EMC), which comprises 6580 thallium-doped CsI crystals. 
These systems are mounted inside a 1.5-T solenoidal
superconducting magnet. 
Muons are identified in the instrumented flux return (IFR), composed
of resistive plate chambers and layers of iron that return the
magnetic flux of the solenoid. 
We use the GEANT4~\cite{geant} software to simulate interactions of
particles traversing the detector, taking into account the varying
accelerator and detector conditions.

\jpsi$\rightarrow e^+e^-$ ($\mu^+\mu^-$) candidates must have a mass between 
$2.95-3.14$ ($3.06-3.14$) \gevcc.
\psitwos candidates are required to have invariant masses 
$3.44 < m_{\epem} < 3.74$ \gevcc or $3.64 < m_{\mumu} < 3.74$ \gevcc. 
Electron candidates are combined with photon candidates in order to
recover some of the energy lost through Bremsstrahlung.
\jpsi candidates and \g candidates with an energy larger than $150
\mev$, are combined to form \chicone candidates, which must satisfy
$350 < m_{\ellell \g} - m_{\ellell} < 450$ \mevcc.
$\piz \ra \gaga$ candidates must satisfy $113 < m_{\gaga} < 153$ \mevcc. 
The energy of each photon has to be greater than $50 \mev$.
$\KS \ra \pip\pim$ candidates are required to satisfy $489 < m_{\pipi} < 507$ \mevcc. 
In addition, the \KS flight distance from the $\psi$ vertex must be larger than three times its uncertainty.
\Kstarz and \Kstarp candidates are required to satisfy 
$796 < m_{\kaon \pi} < 996$ \mevcc and $792 < m_{\kaon \pi} < 992$ \mevcc, 
respectively. 
In addition, due to the presence of a large background of low-energy non-genuine
 \piz's, the cosine of the angle $\theta_{K^*}$ between the \kaon
 momentum and the \B momentum in the \Kstar rest frame has to be less
 than 0.8 for $\Kstar \to\Kpm\piz$.
In events where two $B$'s reconstruct to modes with the same $c\bar c$
and $K$ candidate, one with a \pipm and the other with a \piz, the \B
candidate with a \piz is discarded due to the high background induced
by fake \piz's.

\B candidates, reconstructed by combining $c \bar c$ and \Kstar
candidates, are characterized by two kinematic variables: the difference
between the reconstructed energy of the \B candidate and the beam energy in the 
center-of-mass frame $\DeltaE = E_B^*-\sqrt{s}/2$, and the beam-energy substituted 
mass $\mes \equiv \sqrt{(s/2+{\bf p}_0\cdot {\bf p}_B)^2/E_0^2-{\bf p}_B^{2}}$, 
where subscript $0$ and $B$ correspond to $\Upsilon(4S)$ and the $B$ candidate in the laboratory frame. 
For a correctly reconstructed \B meson, \DeltaE is expected to peak near zero and \mes near the $\B$-meson mass $5.279 \gevcc$. 
The analysis is performed 
in a region of the \mes vs \DeltaE plane defined by $5.2 < \mes < 5.3$ \gevcc and 
$-120 < \DeltaE < 120$ \mev.
The signal region is defined as $\mes > 5.27$ \gevcc and $|\DeltaE|$
smaller than 40 (30) \mev for channels with (without) a \piz. 
For events that have multiple candidates, the candidate having 
the smallest $|\DeltaE|$ is chosen. 
\mes distributions are available in Ref.~\cite{Aubert:2006mg}.

\begin{figure}[h]
\begin{center}
\includegraphics[width=0.9860\linewidth]{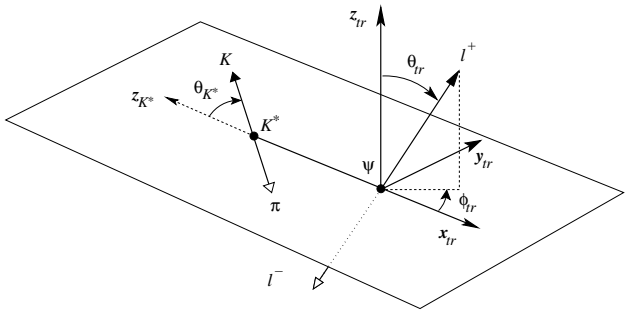}
\caption{\label{fig:heli-trans-frame}Definition of the transversity angles. Details are given in the text.}
\end{center}
\end{figure}

The \B decay amplitudes are measured from the differential decay
distribution, expressed in the transversity basis
\cite{Aubert:2004cp,Aubert:2001pe}, Fig.~\ref{fig:heli-trans-frame},
with conventions detailed in Ref.~\cite{Stephane}.
%
$\theta_{K^*}$ is the helicity angle of the \Kstar decay. It is defined in 
 the rest frame of the \Kstar meson, and is the angle between the kaon 
 and the opposite direction of the $B$ meson in this frame.
$\theta_{\rm tr}$ and $\phi_{\rm tr}$ are defined in the $\psi$
 (\chicone) rest frame and are the polar and azimutal angle of the
 positive lepton (\jpsi daughter of \chicone) , with respect the axis
 defined by:
 \begin{itemize}
 \item $\boldsymbol{x}_{\rm tr}$: opposite direction of the $B$ meson;
 \item $\boldsymbol{y}_{\rm tr}$: perpendicular to $\boldsymbol{x}_{\rm tr}$, in the 
 $(\boldsymbol{x}_{\rm tr},\boldsymbol{p}_{K^*})$ plane, with a direction such that
 $\boldsymbol{p}_{K^*}\cdot \boldsymbol{y}_{\rm tr} > 0$;
 \item $\boldsymbol{z}_{\rm tr}$: to complete the frame, ie: 
 $\boldsymbol{z}_{\rm tr} = \boldsymbol{x}_{\rm tr}\times\boldsymbol{y}_{\rm tr}$.
 \end{itemize}
%
In terms of the transversity angular variables
$\vomega \equiv (\cos{\theta_{K^*}},\cos{\theta_{\rm tr}},\phi_{\rm tr})$, 
the time-integrated differential decay rate for the decay of the $B$ meson is
\begin{equation}
 \label{eqn:g_definition}
 g(\vomega;\boldsymbol{A}) \equiv 
\frac{1}{\Gamma}\frac{{\rm d}^3\Gamma}{{\rm d}\cos\theta_{K^*}{\rm d}\cos\theta_{\rm tr}{\rm d}\phi_{\rm tr}}=
\sum_{k=1}^{6} {\cal A}_k f_k(\vomega),
\end{equation}
where the amplitude coefficients ${\cal A}_i$ and the angular
functions $f_k(\vomega)$, $k=1\cdots 6$ are listed in Table
\ref{tab:diff:decay:rate}.
The $\psi$ decays to two spin-1/2 particles, while the \chicone decays to two vector particles.
The angular
dependencies are therefore different \cite{Stephane}.
\begin{table*}
\caption{\label{tab:diff:decay:rate}
Amplitude coefficients ${\cal A}_k$ and angular functions
$f_k(\vomega)$ that contribute to the differential decay rate.
An overall normalization factor $9/32\pi$ (for $\psi$) and $9/64\pi$
(for $\chicone$) has been omitted. 
In the case of a \Bb decay, the $\Imm$ terms change sign. }
\begin{center}
\begin{ruledtabular}
\begin{tabular}{lcccccccccc} 
$i$  &   ${\cal A}_k$               & $f_k(\vomega)$ for $\psi$ \cite{Aubert:2004cp,Aubert:2001pe} & $f_k(\vomega)$ for \chicone \cite{Stephane}  \\ 
\hline
\noalign{\vskip1pt}
1 & $|A_0|^2                     $ & $2\cos^2\theta_{K^*}\left[1-\sin^2\theta_{\rm tr}\cos^2\phi_{\rm tr}\right] $ & $2 \cos^2{\theta_{K^*}} \left[1+\sin^2{\theta_{\rm tr}}\cos^2{\phi_{\rm tr}}\right]$ \\
2 & $|A_\parallel |^2            $ & $\sin^2\theta_{K^*}\left[1-\sin^2\theta_{\rm tr}\sin^2\phi_{\rm tr}\right] $ & $\sin^2{\theta_{K^*}} \left[1+\sin^2{\theta_{\rm tr}}\sin^2{\phi_{\rm tr}}\right]$ \\
3 & $|A_\perp|^2                 $ & $\sin^2\theta_{K^*}\sin^2\theta_{\rm tr} $ & $\sin^2{\theta_{K^*}}\left[2\cos^2{\theta_{\rm tr}}+ \sin^2{\theta_{\rm tr}}\right]$ \\
4 & $\Imm (A_\parallel ^*A_\perp)$ & $\sin^2\theta_{K^*}\sin2\theta_{\rm tr}\sin\phi_{\rm tr} $ & $-\sin^2{\theta_{K^*}}\sin{2\theta_{\rm tr}}\sin{\phi_{\rm tr}}$ \\
5 & $\Ree (A_\parallel A_0^*)    $ & $-\frac{1}{\sqrt{2}}\sin2\theta_{K^*}\sin^2\theta_{\rm tr}\sin2\phi_{\rm tr} $ & $\frac{1}{\sqrt{2}}\sin{2\theta_{K^*}} \sin^2{\theta_{\rm tr}} \sin{2\phi_{\rm tr}}$ \\
6 & $\Imm (A_\perp A_0^*)        $ & $ \frac{1}{\sqrt{2}}\sin2\theta_{K^*}\sin2\theta_{\rm tr}\cos\phi_{\rm tr} $ & $-\frac{1}{\sqrt{2}}\sin{2\theta_{K^*}}\sin{2\theta_{\rm tr}}\cos{\phi_{\rm tr}}$ \\
\end{tabular}
\end{ruledtabular}
\end{center}
\end{table*}
The symbol $\boldsymbol{A} \equiv (A_0,A_\parallel ,A_\perp)$ denotes the transversity amplitudes for the decay of the \B\ meson, 
and $\boldsymbol{\overline{A}}$ for the $\Bbar$ meson decay. In the absence of direct \CP violation, we
can choose a phase convention in which these amplitudes are related by
$\overline{A}_0 = +A_0 $,
$\overline{A}_\parallel = +A_\parallel $,
$\overline{A}_\perp = -A_\perp $,
so that $A_\perp $ is \CP-odd and $A_0$ and ${A}_\parallel $ are \CP-even.
The phases $\delta_j$ of the amplitudes, where $j= 0, \parallel , \perp$, are
defined by $A_j = |A_j| e^{i \delta_j}$.
Phases are defined relative to $\delta_0=0$. 

We perform an unbinned likelihood fit of the three-dimensional angle probability density function (PDF).
The acceptance of the detector and the efficiency of the event
reconstruction may vary as a function of the transversity angles, in
particular as the angle $\theta_{K^*}$ is strongly correlated with the
momentum of the final kaon and pion.
We use the acceptance correction method developped in Ref.~\cite{Aubert:2004cp}.
The PDF of the observed events, $g_{\rm obs}$, is :
\begin{equation}
\label{eq:gObsDef}
g_{\rm obs}(\vomega;\boldsymbol{A}) = g(\vomega;\boldsymbol{A}) 
 \frac{\varepsilon({\vomega})}{\langle \varepsilon\rangle(\boldsymbol{A})},
\end{equation}
where 

$\varepsilon(\vomega)$ is the angle-dependent acceptance and 
\begin{equation}
\langle \varepsilon\rangle(\boldsymbol{A}) \equiv \int{g(\vomega;\boldsymbol{A})\varepsilon(\vomega){\rm d}\vomega}
\end{equation}
is the average acceptance. 
We take into account the presence of cross-feed from channels
with the same $c\bar{c}$ candidate and a different $K^*$ 
candidate that has (due to isospin symmetry) the same $\boldsymbol{A}$ 
dependence as the signal.
The observed PDF for channel $\jchan$ $(\jchan=\Kpm\pimp, \KS\pipm, \Kpm\piz)$ is then
\begin{eqnarray}
\label{eq:gobs}
g_{\rm obs}^{\jchan}(\vomega;\boldsymbol{A}) &=&
g(\vomega;\boldsymbol{A})
\frac{\varepsilon^{\jchan}(\vomega)}
{\sum_{k=1}^{6}{\cal A}_k(\boldsymbol{A}) \Phi^{\jchan}_k},
\end{eqnarray}
where $\varepsilon^{\jchan}(\vomega)$ is the efficiency,
defined as the ratio between the reconstructed
and generated yield for the process ($B \rightarrow (c\bar c) K^*$, $K^* \rightarrow b$), and we do
not distinguish between correctly reconstructed signal and cross-feed in
the numerator 
\begin{eqnarray}
\label{eqn:effectiveEff}
\varepsilon^{\jchan}(\vomega) &\equiv& \sum_\ichan F_\ichan \varepsilon^{\ichan\to \jchan}(\vomega).
\end{eqnarray}
$\varepsilon^{\ichan\to \jchan}(\vomega)$ is the probability for an event
generated in channel $\ichan$ and with angle $\vomega$ to be detected as an
event in channel $\jchan$.
$F_\ichan, \ichan= \KS\piz, \Kpm\pimp, \Kpm\piz, \KS\pipm$ denotes the fraction
of each channel in the total  branching fraction
$\B\to\ccbar\Kstar$, $\sum_\ichan F_\ichan =1$.
The $\Phi^{\jchan}_k$ are the $f_k(\vomega)$ moments of the
total efficiency $\varepsilon^{\jchan}$, including cross-feed :
\begin{eqnarray}
\Phi^{\jchan}_k &\equiv&
 \sum_\ichan F_\ichan \int{ f_k(\vomega) \varepsilon^{\ichan\to \jchan}(\vomega)
\dd \vomega}.
\label{eqn:coefs:Eff}
\end{eqnarray}

Under the approximations of neglecting the angular resolution for
signal and cross-feed events, and the possible mis-measurement of the
\B flavor such as in events where both daughters in $K^{*0}\to
K^{\pm}\pi^{\mp}$ are mis-identified ($K$-$\pi$ swap), the PDF $g_{\rm
obs}$ can be expressed as in Eq.~(\ref{eq:gObsDef}), and only the
coefficients $\Phi_K^b$ are needed.
The biases induced by these approximations have been estimated with
Monte Carlo (MC) based studies and found to be negligible.

The coefficients $\Phi^{\jchan}_k$ are computed with exclusive
signal MC samples obtained using a full simulation of the experiment
\cite{geant,Evtgen}.
PID efficiencies measured with data control samples
are used to adjust the MC simulation to the observed performance of the detector.
Separate coefficients are used for 
different charges of the final state mesons, in particular to take into account
the charge dependence of the interaction of charged kaons
with matter, and a possible charge asymmetry of the detector.
Writing the expression for the log-likelihood 
$L^{\jchan}(\boldsymbol{A})$
for the PDF 
$g_{\rm obs}^{\jchan}(\vomega_i;\boldsymbol{A})$
for a pure signal sample of $N_{S}$ events, 
the relevant contribution is 
\begin{equation} 
\label{eqn:likelihoodDef}
L^{\jchan}(\boldsymbol{A})=
\sum_{i=1}^{N_{S}}\ln\left( g(\vomega_i;\boldsymbol{A}) \right) - N_{S} \ln\left(\sum_{k}{\cal A}_k(\boldsymbol{A}) \Phi^{\jchan}_k\right), 
\end{equation}
since the remaining term 
$ \sum_{i=1}^{N_{S}}\ln\left(\varepsilon^{\jchan}(\vomega_i)\right)$
does not depend on the amplitudes.

We use a background correction method~\cite{Aubert:2004cp} 
in which background events from a pure
background sample of $N_B$ events are added with a negative weight to
the log-likelihood that is maximized
\begin{equation}
\label{eqn:pseudo-log}
 L^{\prime\jchan}(\boldsymbol{A}) \equiv \sum_{i = 1}^{n_{B}+N_{S}}L(\vomega_i;\boldsymbol{A})
- \frac{\tilde{n}_B}{N_{B}}\sum_{j = 1}^{N_{B}}L(\vomega_j;\boldsymbol{A}),
\end{equation}

\noindent 
where $L(\vomega;\boldsymbol{A})=\ln(g_{\rm obs}^{\jchan}(\vomega;\boldsymbol{A}))$. 
The fit is performed within the \mes signal region. Background events
used here for subtraction are from generic (\BB, \qqbar) MC
samples. $\tilde{n}_B$ is an estimate of the unknown number $n_B$ of
background events that are present in the signal region in the data
sample.

As $L^{\prime\jchan}$ is not a log-likelihood, the uncertainties
yielded by the minimization program \minuit~\cite{minuit} are biased
estimates of the actual uncertainties.
An unbiased estimation of the uncertainties is described and validated
in Appendix~A of Ref.~\cite{Aubert:2004cp}. With this
pseudo-log-likelihood technique, we avoid parametrizing the acceptance
as well as the background angular distributions.
\begin{table*}
\caption{\label{tab-res1}Summary of the measured amplitudes.
For decays to \chicone, as $A_\perp$ is compatible with zero, its phase is not defined.}
\begin{center}
\begin{ruledtabular}
\begin{tabular}{ccccccc}
\noalign{\vskip1pt}
Channel & $|A_0|^2$ & $|A_\parallel|^2$ & $|A_\perp|^2$ & $\delta_\parallel$ & $\delta_\perp$ \\ \hline
$J/{\psi} K^*$ & $0.556\pm0.009\pm0.010$ & $0.211\pm0.010\pm0.006$ & $0.233\pm0.010\pm0.005$ & $-2.93 \pm 0.08 \pm 0.04$ & $2.91 \pm 0.05 \pm 0.03 $ \\ \hline
$\psi(2S) K^*$ & $0.48\pm0.05\pm0.02$ & $0.22\pm0.06\pm0.02$ & $0.30\pm0.06\pm0.02$ & $-2.8 \pm 0.4 \pm 0.1$ & $2.8 \pm 0.3 \pm 0.1 $ \\ \hline
$\chi_{c1} K^*$ & $0.77\pm0.07\pm0.04$ & $0.20\pm0.07\pm0.04$ & $0.03\pm0.04\pm0.02$ & $0.0 \pm 0.3 \pm 0.1$ & -- 
\end{tabular} \end{ruledtabular}
\end{center}
\end{table*}

The measurement is affected by several systematic uncertainties.
The branching fractions used in the cross-feed part of the
acceptance cross section are varied by $\pm 1 \sigma$, and the largest
variation is retained.
The uncertainty induced by the finite size of the MC sample used to
compute the coefficients $\Phi_k^b$ is estimated by the statistical
uncertainty of the angular fit on that MC sample \cite{Aubert:2001pe}.
The uncertainty due to our limited understanding of the PID efficiency is
estimated by using two different methods to correct for the MC-vs-data
differences.
The background uncertainty is obtained by comparing MC and data
shapes of the \mes distributions for the combinatorial component
and by using the corresponding branching errors for the peaking
component.
The uncertainty due to the presence of a $K\pi$ $S$ wave under the
\Kstarone peak is estimated by a fit including it.
The differential decay rate is described by Eqs.~($6$-$9$) of 
 Ref.~\cite{Aubert:2004cp}.

The results are summarized in Table~\ref{tab-res1}.
\begin{figure*}
\begin{center}
\includegraphics[width=0.06\linewidth]{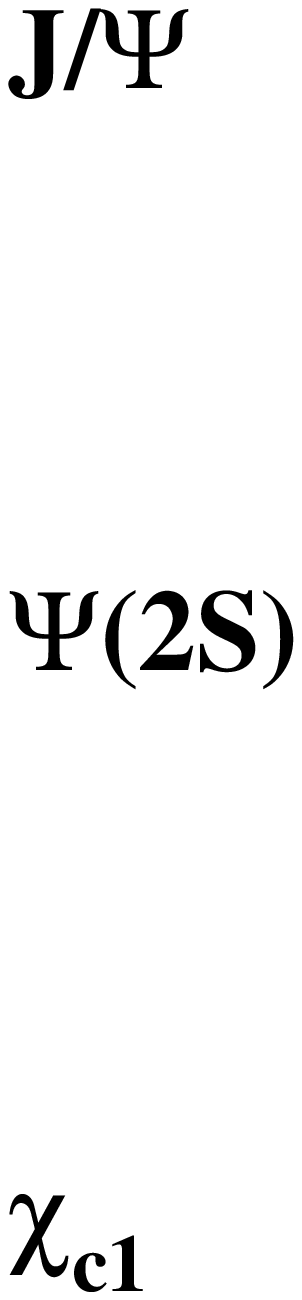}
\includegraphics[width=0.3\linewidth]{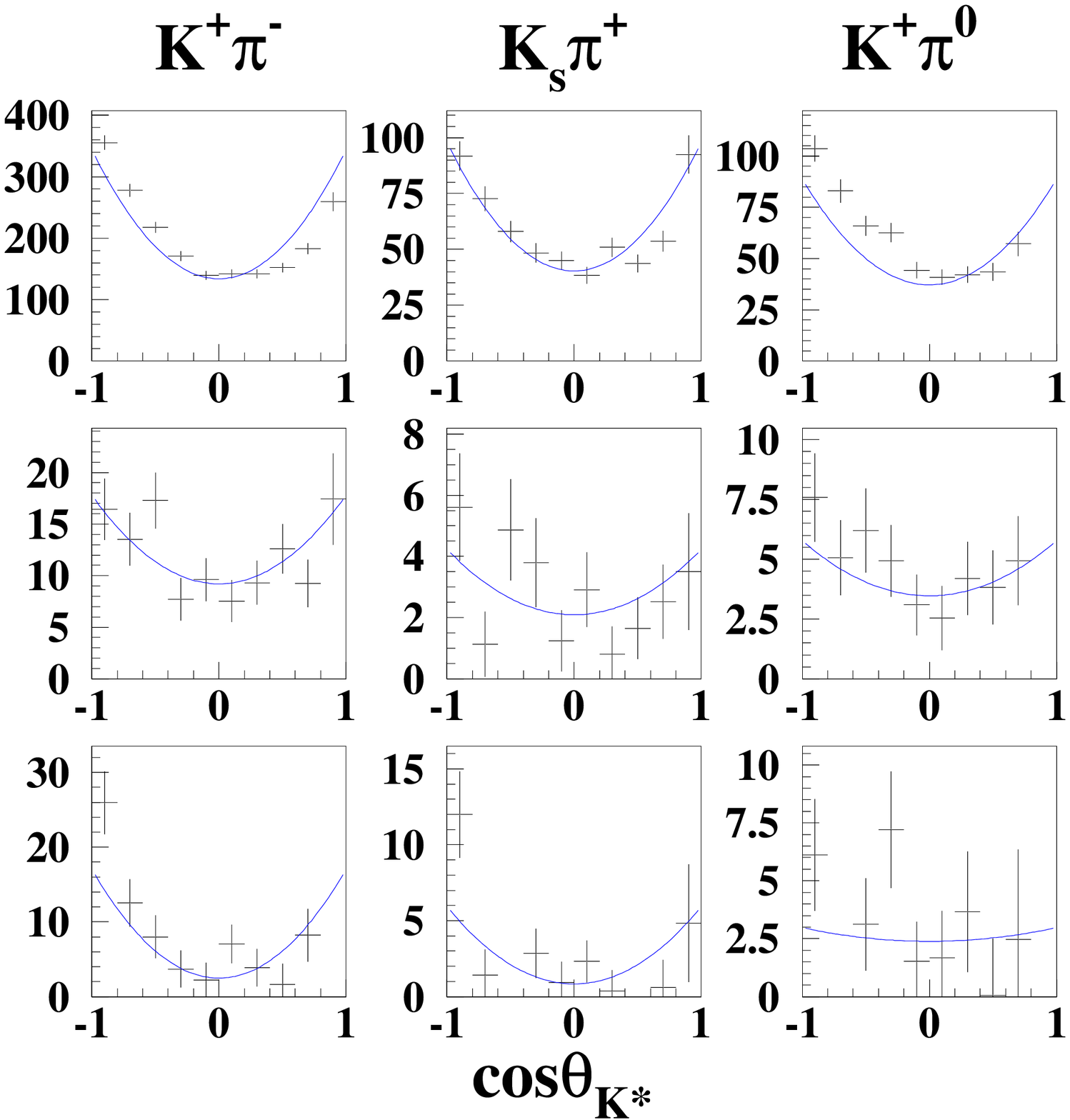}
\hfill
\includegraphics[width=0.3\linewidth]{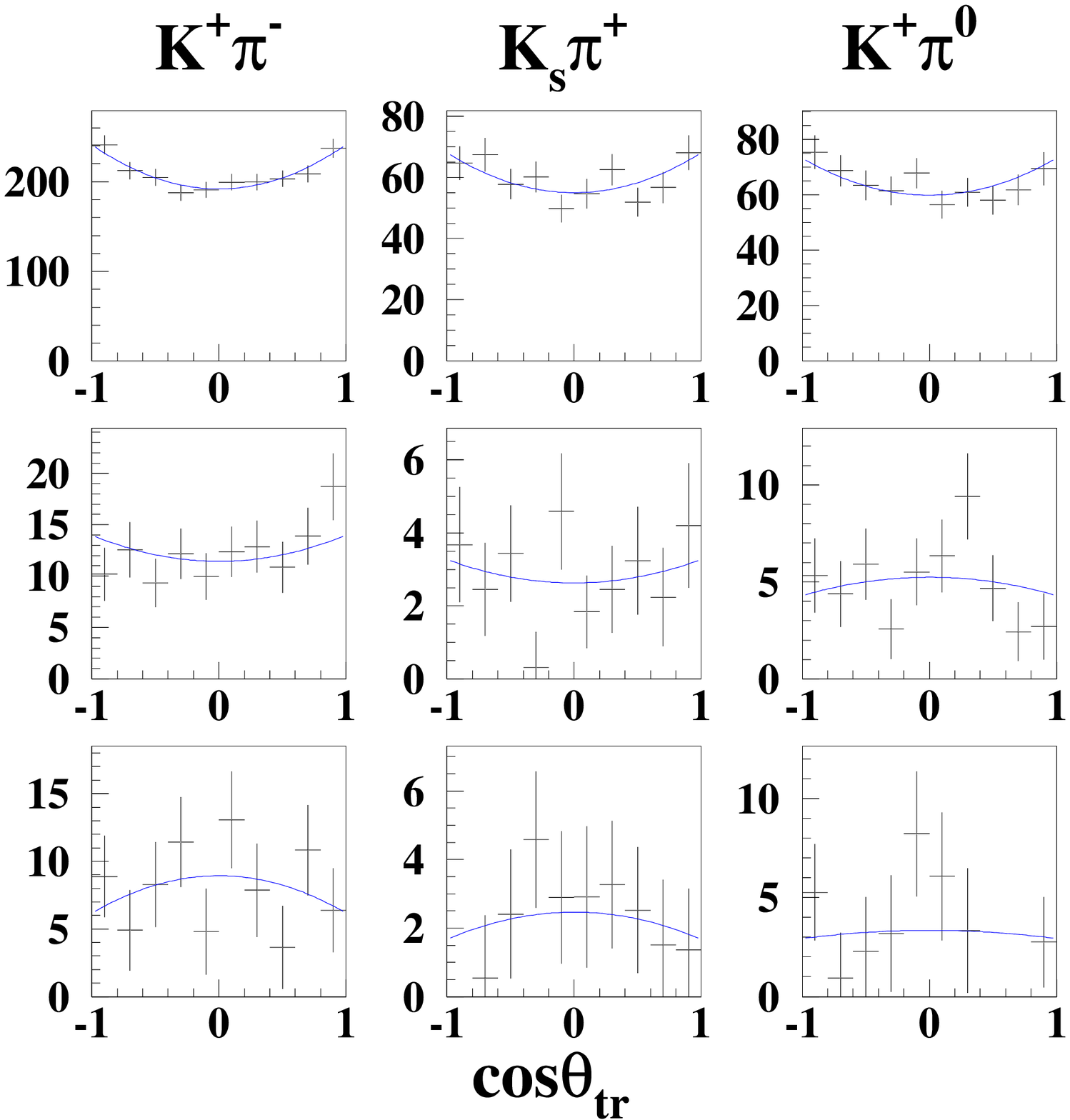}
\hfill
\includegraphics[width=0.3\linewidth]{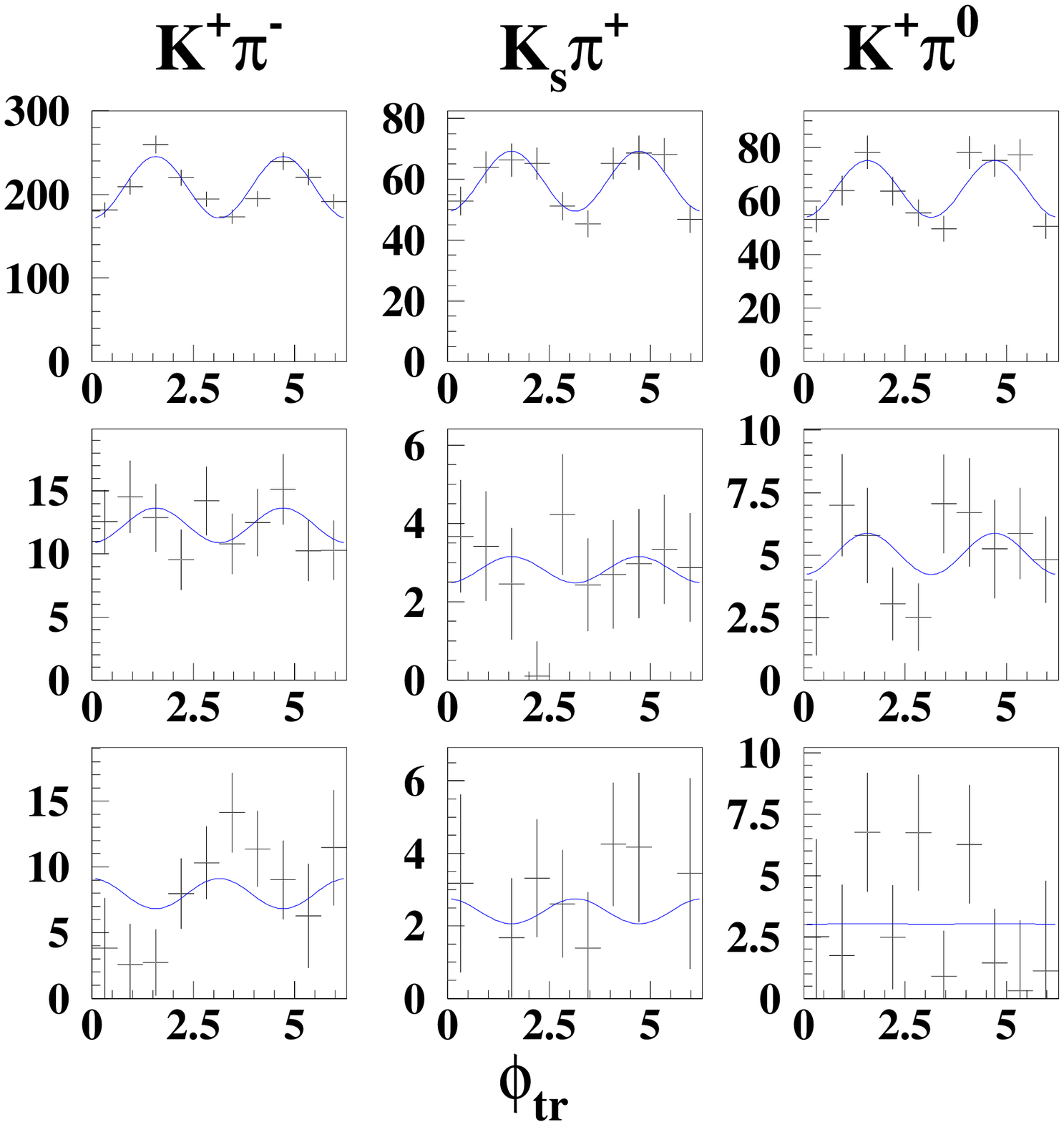}
\caption{\label{fig:resultat}
Angular distributions with PDF from fit overlaid.
The asymmetry of the $\cos\theta_{K^*}$ distributions induced by the S-wave interference is clearly visible.}
\end{center}
\end{figure*}
The values of $|A_0|^2$, $|A_\parallel|^2$, $|A_\perp|^2$ are
 negatively correlated due to the constraint $|A_0|^2 + |A_\parallel|^2 + |A_\perp|^2 =1$. 
In particular, $|A_\parallel|^2$, which would be the
least precisely measured parameter in separate one-dimensional fits, is strongly
anti-correlated with $|A_0|^2$, which would be the best measured.
The one-dimensional (1D) distributions, acceptance-corrected with an
1D Ansatz and background-subtracted, are overlaid with the fit
results and shown on Figure \ref{fig:resultat}.
In contrast with the dedicated method used in the fit, for the plots,
we simply computed the 1D efficiency maps from the distributions of
the accepted events divided by the 1D PDF.
As in lower statistics studies, the $\cos\theta_{K^*}$ forward
backward asymmetry due to the interference with the S wave is clearly
visible.
\begin{table} 
\caption{\label{tab-res:dir} Difference between the interference terms
measured in \B and \Bb decays to \jpsi. }
\begin{center}
\begin{tabular}{c|ccc}
\hline\hline
 & $\delta {\cal A}_4$ & $\delta {\cal A}_6$ \\ \hline
$(K^+ \pi^-)$ & ~ $ 0.002 \pom 0.025 \pom 0.005 $ & $ -0.011 \pom 0.043 \pom 0.016 $ \\
$(K^+ \pi^0)$ & $ -0.017 \pom 0.047 \pom 0.023 $ & $ -0.051 \pom 0.098 \pom 0.064 $ \\
$(\KS \pi^+)$ & $ -0.008 \pom 0.049 \pom 0.011$ & ~ $ 0.075 \pom 0.089 \pom 0.009$ \\
\hline\hline
\end{tabular}
\end{center}
\end{table}

Our measurement of the amplitudes of \B decays to \jpsi are compatible
with, and of better precision than, previous measurements.
A comparison of neutral and charged $B$ decays (not shown) 
yields results consistent with isospin symmetry.
The strong phase difference $\delta_\parallel - \delta_\perp$ is obtained from a fit in which the phase origin is $ \delta_\perp\equiv 0$.
We confirm our previous observation that the strong phase differences
are significantly different from zero, in contrast with what is predicted by factorization.
For $\B\to\jpsi\Kstar$, it amounts to 
$\delta_\parallel - \delta_\perp = 0.45 \pm 0.05 \pm 0.02$.
The presence of direct \CP-violating triple-products in the amplitude
would produce a \B to \Bb difference in the interference terms ${\cal A}_4$ and ${\cal A}_6$:
$\delta {\cal A}_4$ and $\delta {\cal A}_6$. Our results (see Table~\ref{tab-res:dir}), with improved precision relative to Ref.~\cite{belletp}, are consistent with no \CP\ violation.

In summary, we have performed the first three-dimensional analysis of the decays to \psitwos
and \chicone. 
The longitudinal polarization of the decay to \psitwos is
lower than that to \jpsi, 
while the \CP-odd intensity fraction is higher (by 1.4 and 1.0 standard deviations, respectively). This is compatible with the prediction of models of meson decays in the framework of
factorization.
The longitudinal polarization of the decay to \chicone is found to be
larger than that to \jpsi, in contrast with the predictions of
Ref.~\cite{Chen:2005ht}, which include non-factorizable contributions.
The \CP-odd intensity fraction of this decay is 
compatible with zero.
The parallel and longitudinal amplitudes for \chicone seem to be aligned ($|\delta_{\|}-\delta_0|\sim 0$) while for $\psi$ they are anti-aligned ($|\delta_{\|}-\delta_0|\sim\pi$).

We are grateful for the 
extraordinary contributions of our \pep2\ colleagues in
achieving the excellent luminosity and machine conditions
that have made this work possible.
The success of this project also relies critically on the 
expertise and dedication of the computing organizations that 
support \babar.
The collaborating institutions wish to thank 
SLAC for its support and the kind hospitality extended to them. 
This work is supported by the
US Department of Energy
and National Science Foundation, the
Natural Sciences and Engineering Research Council (Canada),
the Commissariat \`a l'Energie Atomique and
Institut National de Physique Nucl\'eaire et de Physique des Particules
(France), the
Bundesministerium f\"ur Bildung und Forschung and
Deutsche Forschungsgemeinschaft
(Germany), the
Istituto Nazionale di Fisica Nucleare (Italy),
the Foundation for Fundamental Research on Matter (The Netherlands),
the Research Council of Norway, the
Ministry of Science and Technology of the Russian Federation, 
Ministerio de Educaci\'on y Ciencia (Spain), and the
Science and Technology Facilities Council (United Kingdom).
Individuals have received support from 
the Marie-Curie IEF program (European Union) and
the A. P. Sloan Foundation.


\end{document}